\documentstyle[prl,preprint,aps]{revtex}
\begin{document}
\draft
%
\title
{Entangled Electronic States in Multiple Quantum-Dot Systems}
\author{S.C. Benjamin and N.F. Johnson}
\vspace{.05in}
%
\address
{Physics Department, Oxford University, Oxford OX1 3PU, England}
%
\maketitle

\begin{abstract}
We present an analytically solvable model of $P$ colinear,
two-dimensional quantum
dots, each containing two electrons. Inter-dot coupling via
the electron-electron interaction
gives rise to sets of entangled ground states.
These ground states
have crystal-like inter-plane correlations and arise
discontinously with increasing magnetic field. Their ranges
and stabilities are
found to depend on dot size ratios, and to increase with $P$.

\end{abstract} \vspace{1.0in}


\narrowtext

Recent experimental and theoretical interest in quantum dot
systems has
opened up the fascinating area
of highly-correlated, few-body quantum phenomena in the
traditionally
large-$N$ field of semiconductor physics. Complex ground state
behavior as a function of magnetic
field
has been predicted for single two-dimensional (2D)
quantum dots
containing $N$ electrons
\cite{maksym}. Even for as few as $N=2$
electrons per 2D dot, ``magic number" ground state transitions
are predicted as a
function of magnetic field as a result
of the competition between (single-electron) confinement
energy and
(many-body) electron-electron interactions
\cite{wagner} \cite{hawrylak}.
Remarkably,
such transitions for $N=2$ have recently been observed
experimentally
\cite{hawrylak}\cite{ashoori}.
Arrays of coupled dots \cite{kempa} \cite{jed}
\cite{chak} \cite{dassarma}
have been attracting increasing attention, partly because
of the possibility of application as ultra-small logic
gates \cite{lent} \cite{qcomputer} \cite{biology}.
Adjacent dots can be coupled by ``optical wiring", i.e.
coupled by the two-body
electron-electron interaction between electrons in adjacent
dots which can be
non-zero even in the {\em absence} of a single-body
tunneling term
\cite{teich}.
Such few-electron coupled 2D systems are also interesting
in that they can represent the
small-$N$ analogs of coupled, parallel 2D electron
gases \cite{dassarma2}.
Given the complex ground state behavior in a single
2D dot as a function
of $B$-field, the coupled 2D dot system offers the
interesting possibility of
competition between electron correlations on the
same dot (i.e. intra-dot interactions
$V_{intra}$) and correlations between adjacent dots
(i.e. inter-dot interactions $V_{inter}$).

This paper predicts the existence of entangled ground states
with crystal-like inter-dot correlations
in multiple quantum dot systems.
These ground states occur discontinously with increasing
magnetic field, being interdispersed with states having
negligible inter-dot correlations. The
ranges and stabilities of the crystal-like states depend on
dot size ratios
and {\em increase} with the number of dots $P$.
Our model is solved analytically and consists of
$P$ colinear 2D quantum dots,
each containing two electrons but not necessarily identical
in size.
The model considers a sufficient number of electrons
as to contain both inter {\em and}
intra-dot electron-electron interactions, and yet still
admit analytic solutions. These analytic solutions implicitly
include mixing with all Landau levels.

Figure 1 shows our system for a pair of dots ($P=2$).
The dots are arranged vertically with separation $s$.
This vertical geometry is of specific experimental interest
given
the possibility of fabrication via etching of a multiple quantum
well
structure
(see Ref. \cite{pepper} for $P=2$).
Following several single-dot
studies \cite{maksym}, we model each of the $P$ dots by
2D (xy plane) parabolic
potentials with a perpendicular magnetic field B (z direction)
of sufficient
strength to spin-polarize the electrons.
Electrostatic confinement in the
z-direction is taken to be sufficiently strong that the
electrons are frozen in the lowest z sub-band. We take the
electron-electron interaction potential to be of
inverse-square form,
$\frac{\beta}{r^2}$ where $\beta$ is a positive parameter; this
interaction for a single layer gives quantitatively similar
results to the bare Coulomb interaction \cite{quiroga}
\cite{kinaret}.
Inter-dot
tunneling is assumed to be negligible.
The Hamiltonian for the system of $P$ dots is (with a
symmetric gauge)
$H=H_0 + V$ with
\begin{equation}
H_0 =\sum_{\alpha=1,2,\dots P
;i=1,2} ({{\bf p}_{\alpha,i}^2\over {2m^*}} +
{1\over 2}m^*\omega_{\alpha}^2(B)|{\bf r}_{\alpha,i}|^2 +
{\omega_c\over
2}L_{\alpha,i})\ ;
\end{equation}
${\bf p}_{\alpha,i}$ and ${{\bf r}_{\alpha,i}}$ are
the momentum and position of electron $i$ in dot $\alpha$.
Each electron has effective mass $m^*$ and z-component of
angular momentum
${L_{\alpha,i}}$. The
cyclotron frequency is $\omega_c$ and
$\omega_{\alpha}^2(B)={\omega_{0\alpha}^2+\omega_c^2/4}$. The
electrostatic confining potential $\omega_{0\alpha}$ is in
general different for each
dot (N.B. the dot
`size' $\sim\omega^{-\frac{1}{2}}_{0\alpha}$).
The dominant inter-dot coupling is due to
interactions between electrons on adjacent dots; we therefore
take
$V=V_{intra}+V_{inter}$ where
\begin{equation}
V_{intra}=
\sum_{\alpha=1,2,\dots P}{\beta\over {|{\bf r}_{\alpha,1}-
{\bf r}_{\alpha,2}|^2}}
\end{equation}
and
\begin{equation}
V_{inter}=\sum_{\alpha=2,3,\dots P}{\sum_{i,j}{\beta\over {|{\bf
r}_{\alpha,i}-{\bf r}_{(\alpha-1),j}|^2 + s^2}}}\ . \end{equation}
\noindent For $V_{inter}\rightarrow 0$,
the exact eigenstates of $H$
are products of $P$ single-dot states \cite{quiroga}.
For finite $V_{inter}$, no exact analytic solutions are known.
Our approach is to Taylor expand $V_{inter}$ under the assumption
that $|{\bf r}_{\alpha,i}-{\bf r}_{(\alpha-1),j}|^2<s^2$. Having
solved the resulting problem,
we can check from the analytic expressions for
(e.g.) ${ r^2\over
s^2}$ that any particular set of dot parameters is consistent with
this assumption.
Hence
\begin{equation}
V_{inter}={\beta\over s^2}\sum_{\alpha=2,3,\dots P}\sum_{i,j}
{\sum_{k=0}^{\infty}{(-1)^k({|{\bf r}_{\alpha,i}-
{\bf r}_{(\alpha- 1),j}|^2
\over{s^2}})^{k}}}
\end{equation}
Exact analytic solution of $H$ is now possible including terms
of order $\beta r^2/s^4$
in $V_{inter}$;
(analytic) perturbation theory
is then employed for the $\beta r^4/s^6$ terms.

We now discuss the method explicitly for $P=2$.
Employing an orthogonal transformation, with coefficients
depending on the relative dot sizes $\omega_1$ and $\omega_2$,
allows exact solution of $H$ including terms of order
$\beta r^2/s^4$ in $V_{inter}$.
The only non-trivial contributions to $H$ to this order are
those which depend on the relative positions of the electrons within
each dot. The corresponding eigenstates contain quantum
numbers $m_\alpha$ (the relative angular momentum between the
two electrons in dot $\alpha$).
The ground and low-lying
states of the $P=2$ system have all other quantum numbers zero;
these states can be labelled $|m_1,m_2\rangle$ (signifying the
direct product of $|m_1\rangle$ and $|m_2\rangle$)
and are used as
a basis
for the $\beta r^4/s^6$ perturbation.
Figure 2(a) shows the ground state transitions
as a function of
$B$-field in the limit $s\rightarrow\infty$ (i.e.
$V_{inter}\rightarrow 0$) for the case
of equal dot sizes ($\omega_1=\omega_2$).
As a demonstration of the (lack of) inter-dot
correlation, the insets show the charge density
in a given dot with the
electrons in the other dot `fixed' opposite one another
at the crosses.
The plots are angularly symmetric
(i.e. negligible inter-dot correlation).
The radial
localization increases as the ground state $m_\alpha$
increases
(i.e. as $B\rightarrow\infty$, each dot
approaches its own classical limit).
Figure 2(b) shows the corresponding diagram for
finite $s$.
The main feature is that new entangled  \cite{entangled}
states arise as a
result of finite $V_{inter}$ (i.e. including
terms to order $\beta r^4/s^6$). The density plots show the strong
inter-dot correlation which
characterises the new entangled ground states.

These entangled states
can be easily understood for equal dot sizes.
First note that in the exactly
solved (i.e. order $\beta r^2/s^4$) system the states
$|m_a,m_b\rangle$ and $|m_b,m_a\rangle$
are degenerate.
 Secondly, the perturbation (i.e. order
$\beta r^4/s^6$ terms) only mixes $|m_a,m_b\rangle$
and $|m_c,m_d\rangle$ if $m_a-m_c=m_d-m_b=0$ or
$\pm2$. The mixed (i.e.
entangled) states are therefore
$(|m+2,m\rangle\pm|m,m+2\rangle)$ which we
write as $|m+2,m\rangle_+$ and $|m+2,m\rangle_-$.
It is remarkable that as $B$ increases the ground state
switches back and
forth between `pure' states ($m_1=m_2$) and entangled
states (N.B. the $|m,m\rangle$
states are now not strictly
`pure' because of a very small
non-degenerate perturbation mixing).
For large $B$ ($B>10T$), the entangled states
prevent the pure states from becoming ground states.
As $B\rightarrow\infty$, the
classical limit is reached of four point charges situated
at the corners of a square
when projected onto the $xy$-plane. An alternative view of
the formation of these
entangled states is as a resonance phenomenon; for example,
the
state $|3,1\rangle$ can be thought of as continually
exchanging energy with
$|1,3\rangle$ via virtual photons (i.e. via the
electromagnetic field
representing the electron-electron interaction).
This is equivalent to the resonant Forster
process which is well-known as an energy transfer
mechanism for biological
molecules and proteins. In excitonic language, the
states $|3,1\rangle$ and $|1,3\rangle$ correspond to the
$|1,1\rangle$ vacuum plus an electron-hole excitation
(exciton) of angular momentum $2$ on dots $1$ and $2$
respectively; the formation of the
entangled state $|3,1\rangle_-$
represents the resonance between these two adjacent excitons.

Figures 3(i)a and 3(i)b
show in detail the region near the
$|1,1\rangle$ and $|3,3\rangle$ crossover discussed above.
Figures 3(ii)a and 3(ii)b show the corresponding region for
the case of unequal dot
sizes ($\omega_1\not=\omega_2$). In this case $|m_a,m_b\rangle$
and $|m_b,m_a\rangle$ are no longer degenerate. However by
keeping $|\omega_1- \omega_2|<<\omega_1$, we can arrange
that the two states
are energetically much closer to one
another than to any other state with which they can mix.
Under these conditions we can again solve the problem
analytically;
the entangled states
are $(p|m+2,m\rangle\pm q|m,m+2\rangle)$ where $p$,$q$ are
unequal and depend on the
energies of the component states.
Figure 3(ii)b shows that the stability of the entangled
ground states, i.e. the gap between the entangled ground
states and the other
competing states, is greater than the corresponding gap
in Fig. 3(i)b. The gap scales approximately as $\sqrt{\epsilon^2+
(\Delta E)^2}$ where
$\epsilon$ is the
gap in the equal dot system (Fig.3(i)b) and $\Delta E$
is the energy difference
between component states (i.e. the separation of the
dashed lines in Fig. 3(ii)a). The
results of Figs. 3(i) and 3(ii) are analogous to
linear and quadratic Stark effects;
in both Figs. 3(i) and 3(ii) we essentially have two-level
systems with the
former being degenerate while the latter is
non-degenerate.

Figures 3(iii)a and 3(iii)b consider the case of
$P=3$ equal-size dots.
Exact analytic treatment to order $\beta r^2/s^4$ proceeds
as above. The ground and low-lying
states are now characterised by three non-zero quantum
numbers describing the relative
angular momentum between electrons on each of the three dots;
we label these states $|m_1,m_2,m_3\rangle$.
Just as for $P=2$ with $\omega_1=\omega_2$ to
this order (i.e. $\beta r^2/s^4$) we find degeneracies:
$|m_a,m_b,m_c\rangle$ is degenerate with
$|m_c,m_b,m_a\rangle$. In the limit of large
$s$ all other states formed by
permuting $m_a,m_b,m_c$ would also be degenerate.
Having exactly solved for $P=3$ to order $\beta r^2/s^4$,
we again turn to perturbation theory
to treat the terms of order $\beta r^4/s^6$
(which we will denote by $h$).
The non-zero matrix elements  of $h$ in the nearly degenerate
subspace are found to be of
the form $\langle m_a+X,m_b+Y,m_c+Z|h|m_a,m_b,m_c\rangle$
where $\{X,Y,Z\}$ is any
permutation of $\{-2,0,2\}$. For example,
$|3,1,1\rangle$ mixes with
$|1,3,1\rangle$ and $|1,1,3\rangle$ yielding
three new mixed (i.e. entangled) states.
These entangled states then compete with each
other to become ground states for finite $s$
(see Fig. 3(iii)b).
The entangled
states are the $P=3$ dot analogs of the
inter-dot correlated, crystal-like
states for $P=2$.

As the number of dots $P$ increases,
the gain in energy of the entangled
(i.e. inter-dot
correlated) states at finite $s$
actually {\em
increases} as compared
to the $s\rightarrow\infty$ limit. This increase is non-linear
with $P$.
Note that we can think of the earlier cases of
$P=2$ and $P=3$ as a `diatomic molecule' and
`triatomic molecule' respectively.
Consider the
`polyatomic molecule' with $P$ identical
`atoms' (dots) all in the $m=1$ state.
The state $m=3$, for example,  can be created on
any one of these dots yielding
$P$ degenerate `tight-binding' combinations or
`molecular states'
in the limit $s\rightarrow\infty$.
The degeneracy of these states is
broken for finite $s$ by
the coupling
between adjacent dots.
In contrast to an actual tight-binding molecule where
it is the
single-body tunneling term that couples atoms,
here the coupling is via the two-body
interaction $V_{inter}$.
The states represent
`Frenkel excitons'; the exciton (i.e. $m=3$ state)
on a given dot
can transfer its energy resonantly to
all members of the chain.
For large $P$, the analysis is simplified considerably
by introducing periodic boundary conditions.
This removes `end-effects', introducing
translational symmetry into the problem and allowing application
of Bloch's theorem.
Consider a system of $P$ dots with
a lattice structure;
if $\omega_i=\omega$ for $i=1,\dots P$ then this is a `monatomic
crystal'. If alternating $\omega$'s differ, we have a `diatomic crystal'.
The entangled states are now travelling wave excitations and
yield an exciton band structure.

In summary, we have presented details of entangled
ground states arising in
colinear, multiple dot structures.
The tendency for formation of such states
{\em increases} with the number of dots.

\vskip\baselineskip
This work was supported by
an EPSRC Studentship (S.B.) and the Nuffield
Foundation (N.F.J.) .

\newpage
\centerline{\bf Figure Captions}
\noindent Figure 1:

Schematic illustration of the $P=2$ dot system.
Each dot contains two electrons.
\bigskip

\noindent Figure 2:

\noindent (a) Energies (in Kelvin relative to an arbitrary zero)
of
low-lying eigenstates $|m_1,m_2\rangle$ as a function
of $B$
for $P=2$ dot system in limit $s\rightarrow\infty$.
Lowest curve at a given $B$
corresponds to the ground state. From left to right
the solid lines represent
successive ground states $|1,1\rangle$, $|3,3\rangle$
and $|5,5\rangle$;
dashed lines correspond to the degenerate pairs
$(|3,1\rangle$, $|1,3\rangle)$
and $(|5,3\rangle$, $|3,5\rangle)$.
Contour plots are ground state charge densities
in a given dot
with electrons in the other dot fixed at the crosses.

\noindent (b) As (a), but with {\it finite} $s$.
The degenerate pairs from (a)
split to form entangled states;
the dashed lines to the left represent new
states $|3,1\rangle_{+}$ and $|3,1\rangle_{-}$;
those to the right are $|5,3\rangle_{+}$ and
$|5,3\rangle_{-}$.
States $|3,1\rangle_{-}$ \& $|5,3\rangle_{-}$
become ground states;
their charge densities are shown.

\bigskip

\noindent Figure 3:

\noindent Energies (in Kelvin relative to an
arbitrary zero)
of low-lying states as a function of $B$ for
three different systems. (i) $P=2$ dots of equal
sizes. Parts (a) and (b) are magnifications
from Figs. 2(a) and 2(b) respectively.
(ii) $P=2$ dots of unequal size.
(iii) $P=3$ dots of equal size;
in (a) each dashed line is triply degenerate while
in (b) only the lowest two of the six entangled
states are shown.


\begin{thebibliography}{99}
\bibitem{biology}
Such
weakly-coupled arrays of confined few-electron
systems are not uncommon
in various naturally occurring biological systems.
\bibitem{entangled}
Entangled states cannot be written as a
direct product ($|m_1,m_2\rangle$).
\end{thebibliography}
\end{document}